\documentclass[11pt]{article}
\usepackage{mathrsfs}
\usepackage{amsfonts}
\usepackage{amsmath,amssymb}
\usepackage{graphicx}
\usepackage{caption2}
\usepackage{epsfig}
\usepackage{subfigure}
\usepackage[dvips]{color}

\def\dref#1{(\ref{#1})}
\def\rm{\mathrm}
\begin{document}

\begin{center}
{\LARGE\bf{Distributed Consensus of Linear Multi-Agent\\
\vspace*{5pt}
 Systems with Adaptive Dynamic Protocols} \footnote[1] {\small Zhongkui Li and
Xiangdong Liu are with the School of Automation, Beijing Institute
of Technology, Beijing 100081, P. R. China (e-mails:
zhongkli@gmail.com,xdliu@bit.edu.cn). 
Wei Ren is with the
Department of Electrical Engineering, University of California,
Riverside, CA, 92521, USA (e-mail: ren@ee.ucr.edu). Lihua Xie is
with the School of Electrical and Electronic Engineering, Nanyang
Technological University, 639798, Singapore (e-mail:
elhxieg@ntu.edu.sg).}}
\end{center}
\vskip 0.3cm \centerline{Zhongkui Li, Xiangdong Liu, Wei
Ren, Lihua Xie}

\vskip 0.3cm
\centerline{\today}

\vskip 1cm

{\noindent \small {\bf  Abstract}:
This paper considers the distributed consensus problem
of multi-agent systems with general continuous-time
linear dynamics.
Two distributed adaptive dynamic consensus protocols are proposed,
based on the relative
output information of neighboring agents. One protocol assigns
an adaptive coupling weight to each edge in the communication graph
while
the other uses an adaptive coupling weight for each node.
These two adaptive protocols are designed to ensure that
consensus is reached in a fully distributed fashion for
any undirected connected communication graphs without using any
global information.
A sufficient condition for the
existence of these adaptive protocols is
that each agent is stabilizable and detectable.
The cases with leader-follower and switching communication
graphs are also studied.

\vskip 0.2cm

{\noindent \bf Keywords}:  Multi-agent system, consensus,
cooperative control, adaptive control, relative output.}

\vskip 0.6cm
\section{Introduction}

Consensus is an important problem in the area of cooperative
control of multi-agent systems. The main idea of consensus
is to develop distributed control policies that enable a group of
agents to reach an agreement on certain quantities of interest.
Due to its potential
applications in broad areas such as spacecraft formation flying and sensor networks,
the consensus problem has been extensively studied by numerous
researchers from various perspectives;
see \cite{jadbabaie2003coordination,
olfati-saber2004consensus,ren2005consensus,olfati-saber2007consensus,ren2007information,
ren2008consensus,li2010distributed}
and references therein.
Specifically,
a general framework of the
consensus problem for networks of integrators with fixed or
switching topologies is proposed in \cite{olfati-saber2004consensus}.
The controllability of leader-follower multi-agent systems is considered
in \cite{rahmani2008controllability} from a graph-theoretic perspective.
Distributed tracking control for multi-agent consensus with an active leader is
addressed in \cite{hong2008distributed,hu2010distributed} by using neighbor-based
state estimators. Consensus
of networks of double- and high-order integrators is studied
in \cite{ren2008consensus,ren2007high-order,jiang2009consensus}.
Consensus algorithms are designed in \cite{li2010distributed,carli2009quantized} for a group
of agents with quantized communication links and limited data rate.
In most existing studies on consensus,
the agent dynamics are assumed to be first-,
second-, or high-order integrators, which might be restrictive in
many cases.

This paper considers the distributed consensus problem of multi-agent systems
with general continuous-time linear dynamics. Previous works along
this line include
\cite{li2010consensus,li2011consensusSCL,tuna2009conditions,seo2009consensus,scardovi2009synchronization,zhang2011optimal,ma2010necessary}.
One common feature in \cite{li2010consensus,li2011consensusSCL,seo2009consensus,zhang2011optimal,ma2010necessary}
is that at least the smallest nonzero eigenvalue of the Laplacian matrix associated with the
communication graph is required to be known for the consensus protocol design.
However, the smallest nonzero eigenvalue of the Laplacian matrix
is global information in the sense that
each agent has to know the entire
communication graph to compute it. Therefore, the
consensus protocols given in
\cite{li2010consensus,li2011consensusSCL,seo2009consensus,zhang2011optimal,ma2010necessary}
cannot be implemented by the agents in a
fully distributed fashion, i.e., using only the local information of
its own and neighbors. To overcome this limitation,
an adaptive static consensus protocol is proposed in \cite{li2011adaptive},
which is motivated by
the adaptive strategies for synchronization of
complex networks in \cite{de2008synchronization}. Similar
adaptive schemes are presented to achieve second-order
consensus with inherent nonlinear dynamics in \cite{yu2011distributed}.
Note that the protocols in \cite{li2011adaptive,de2008synchronization,yu2011distributed}
rely on
the relative states of neighboring agents, which
however might not be available in many circumstances.


In this paper, we extend \cite{li2011adaptive,yu2011distributed}
to investigate
the case where the relative outputs,
rather than the relative states, of neighboring agents
are accessible.
Two novel distributed adaptive dynamic consensus protocols
are proposed, namely, one protocol assigns
an adaptive coupling weight to each edge in the communication graph
while
the other uses an adaptive coupling weight for each node.
These two adaptive protocols are designed to ensure that
consensus is reached in a fully distributed fashion for
any undirected connected communication graph without using any
global information.
A sufficient condition for the
existence of these adaptive protocols is
that each agent is stabilizable and detectable.
The cases with leader-follower and switching graphs are also studied.
It is shown that the consensus protocol with an adaptive coupling weight
for each edge is applicable to arbitrary switching
connected communication graphs.
It is worth mentioning that the consensus
protocols in
\cite{tuna2009conditions,scardovi2009synchronization} do not need
any global information either. However, in \cite{tuna2009conditions}
the protocol is based on the relative states
of neighboring agents and
the agent dynamics
are restricted to be
neutrally stable.
In \cite{scardovi2009synchronization},
the eigenvalues of the state matrix of each agent
are assumed to lie in the closed
left-half plane.
Furthermore, the
dimension of the protocol in \cite{scardovi2009synchronization}
is higher than that of the consensus protocol with an adaptive coupling weight
for each node in the current paper.

The rest of this paper is organized as follows. Some useful results
of the graph theory are reviewed in Section 2.
The consensus problems under the proposed
two distributed adaptive protocols are investigated in Section 3.
The consensus problem with leader-follower and switching
communication graphs are studied, respectively, in Sections 4 and 5.
A simulation example is presented in Section 6
to illustrate the analytical results.
Section 7 concludes the paper.


\section{Notation and Graph Theory}

Let $\mathbf{R}^{n\times n}$ be the set of $n\times n$ real
matrices. The superscript $T$ means the transpose for real matrices.
$I_N$ represents the identity matrix of dimension $N$. Matrices, if
not explicitly stated, are assumed to have compatible dimensions.
Denote by $\mathbf{1}$ the column vector with all entries equal to
one. The matrix inequality
$A>B$ (respectively, $A\geq B$) means that $A-B$ is positive
definite (respectively, positive semi-definite). $A\otimes B$
denotes the Kronecker product of matrices $A$ and $B$. ${\rm{diag}}(A_1,\cdots,A_N)$
represents a block-diagonal matrix with matrices $A_i,i=1,\cdots,N,$
on its diagonal.
A matrix is Hurwitz if all of its eigenvalues have negative real parts.

A directed graph $\mathcal {G}$ is a pair $(\mathcal {V}, \mathcal
{E})$, where $\mathcal {V}=\{v_1,\cdots,v_N\}$ is a nonempty finite
set of nodes and $\mathcal {E}\subseteq\mathcal {V}\times\mathcal
{V}$ is a set of edges, in which an edge is represented by an
ordered pair of distinct nodes. For an edge $(v_i,v_j)$, node $v_i$
is called the parent node, node $v_j$ the child node, and $v_i$ is a
neighbor of $v_j$. A graph with the property that
$(v_i,v_j)\in\mathcal {E}$ implies $(v_j, v_i)\in\mathcal {E}$ for
any $v_i,v_j\in\mathcal {V}$ is
said to be undirected. A path from node $v_{i_1}$ to node $v_{i_l}$
is a sequence of ordered edges of the form $(v_{i_k}, v_{i_{k+1}})$,
$k=1,\cdots,l-1$.
An undirected graph is connected if there exists a path between every pair of distinct nodes,
otherwise is disconnected.
A directed graph contains a directed spanning tree
if there exists a node called the root, which has no parent node,
such that the node has directed paths to all other nodes in the
graph.

The adjacency matrix $\mathcal {A}=[a_{ij}]\in\mathbf{R}^{N\times
N}$ associated with the directed graph $\mathcal {G}$ is defined by
$a_{ii}=0$, $a_{ij}=1$ if $(j,i)\in\mathcal {E}$ and $a_{ij}=0$
otherwise. The Laplacian matrix $\mathcal {L}=[\mathcal
{L}_{ij}]\in\mathbf{R}^{N\times N}$ is defined as $\mathcal
{L}_{ii}=\sum_{j\neq i}a_{ij}$ and $\mathcal {L}_{ij}=-a_{ij}$,
$i\neq j$. For undirected graphs, both $\mathcal {A}$ and
$\mathcal {L}$ are symmetric.

{\bf Lemma 1} \cite{godsil2001algebraic,olfati-saber2004consensus,ren2005consensus}.

(1) Zero is an
eigenvalue of $\mathcal {L}$ with $\mathbf{1}$ as a corresponding
right eigenvector, and all nonzero eigenvalues have positive real
parts. Furthermore, zero is a simple eigenvalue of $\mathcal {L}$ if
and only if the graph $\mathcal {G}$ has a directed spanning tree.

(2) For an undirected graph $\mathcal {G}$, the smallest nonzero
eigenvalue $\lambda_2(\mathcal {L})$ of the Laplacian matrix $\mathcal {L}$ satisfies
$\lambda_2(\mathcal {L})=\underset{x\neq0,\mathbf{1}^Tx=0}{\min}\frac{x^T\mathcal {L}x}{x^Tx}.$

\section{Consensus with Undirected Communication Graphs}

In this section,
we assume that the communication graph among the agents, denoted by $\mathcal{G}$,
is undirected. Consider a group of $N$ identical agents with general
linear dynamics. The dynamics of the $i$-th agent
are described by
\begin{equation}\label{1c}
\begin{aligned}
    \dot{x}_i &=Ax_i+Bu_i,\\
    y_i & = Cx_i,\quad i=1,\cdots,N,
\end{aligned}
\end{equation}
where $x_i\in\mathbf{R}^n$ is the state,
$u_i\in\mathbf{R}^{p}$ the control input, $y_i\in\mathbf{R}^{q}$
the measured output, and $A$, $B$, $C$ are constant matrices with
compatible dimensions.

In order to achieve consensus for the agents in \dref{1c}, a variety of static
and dynamic consensus protocols have been proposed in, e.g., \cite{li2010consensus,li2011consensusSCL,tuna2009conditions,seo2009consensus,scardovi2009synchronization,zhang2011optimal,ma2010necessary}.
One common feature in \cite{li2010consensus,li2011consensusSCL,seo2009consensus,zhang2011optimal,ma2010necessary}
is that at least the smallest nonzero eigenvalue $\lambda_2$ of
the Laplacian matrix associated with $\mathcal {G}$
is required to be known for the consensus protocol design.
However, $\lambda_2$ is global information in the sense that
each agent has to know the entire
graph $\mathcal {G}$ to compute it. Therefore, the
consensus protocols given in \cite{li2010consensus,li2011consensusSCL,seo2009consensus,zhang2011optimal,ma2010necessary}
cannot be implemented by the agents in a
fully distributed fashion, i.e., using only the local information of
its own and neighbors. To overcome this limitation,
an adaptive static consensus protocol is proposed in
\cite{li2011adaptive}, which is motivated by
the adaptive strategies for synchronization of
complex networks in \cite{de2008synchronization}. 
Note that the protocols in \cite{li2011adaptive,de2008synchronization}
are based on
the relative states of neighboring agents, which
however might not be available in many circumstances.

In this paper, we extend to
investigate the case where
each agent knows the relative outputs, rather than the relative states,
of its neighbors with respect to itself.
Two novel distributed adaptive dynamic consensus protocols are proposed.
The first adaptive
consensus protocol dynamically updates the coupling weight
for each edge (i.e., the communication links between neighboring agents),
which is given by
\begin{equation}\label{clc3}
\begin{aligned}
\dot{v}_i &=(A+BF)v_i+L\sum_{j=1}^Nc_{ij}a_{ij}\left[C(v_i-v_j)-(y_i-y_j)\right],\\
\dot{c}_{ij}&
=\kappa_{ij}a_{ij}\begin{bmatrix}y_i-y_j \\ C(v_i-v_j)\end{bmatrix}^T\Gamma
\begin{bmatrix}y_i-y_j \\ C(v_i-v_j)\end{bmatrix},\\
u_i &
=Fv_i,\quad i=1,\cdots,N,
\end{aligned}
\end{equation}
where $v_i\in\mathbf{R}^{n}$ is the protocol state,
$i=1,\cdots,N$, $a_{ij}$ is the
$(i,j)$-th entry of the adjacency matrix $\mathcal {A}$ associated with
$\mathcal {G}$, $c_{ij}(t)$
denotes the time-varying coupling weight for the edge $(i,j)$ 
with $c_{ij}(0)=c_{ji}(0)$,
$\kappa_{ij}=\kappa_{ji}$ are positive constants,
and $L\in\mathbf{R}^{q\times n}$, $F\in\mathbf{R}^{p\times n}$,
and $\Gamma\in\mathbf{R}^{2q\times 2q}$
are gain matrices to be determined.

The second adaptive consensus protocol assigns an adaptive coupling
weight to each node (i.e., agent), described by
\begin{equation}\label{clc4}
\begin{aligned}
\dot{\tilde{v}}_i &=(A+BF)\tilde{v}_i+d_iL\sum_{j=1}^Na_{ij}\left[C(\tilde{v}_i-\tilde{v}_j)-(y_i-y_j)\right],\\
\dot{d}_{i}& =\tau_{i}\left(\sum_{j=1}^Na_{ij}\begin{bmatrix}y_i-y_j \\
C(\tilde{v}_i-\tilde{v}_j)\end{bmatrix}^T\right)\Gamma
\left(\sum_{j=1}^Na_{ij}\begin{bmatrix}y_i-y_j \\ C(\tilde{v}_i-\tilde{v}_j)\end{bmatrix}\right),\\
u_i &
=F\tilde{v}_i,\quad i=1,\cdots,N,
\end{aligned}
\end{equation}
where $\tilde{v}_i\in\mathbf{R}^{n}$ is the protocol state,
$i=1,2,\cdots,N$, $d_{i}(t)$
denotes the coupling weight for agent $i$,
$\tau_{i}$ are positive constants, and the rest
of the variables are defined as in \dref{clc3}.

The objective in this section is to find proper gain matrices in
\dref{clc3} and \dref{clc4} such that
the $N$ agents in \dref{1c} achieve consensus in the
sense of $\lim_{t\rightarrow \infty}\|x_i(t)- x_j(t)\|=0$,
$\forall\,i,j=1,\cdots,N.$

\subsection{Consensus Under Adaptive Protocol \dref{clc3}}

In this section, we study the consensus problem of the agents in \dref{1c}
under the adaptive protocol \dref{clc3}.
Let $z_i=[x_i^T,v_i^T]^T$, $e_i=z_i-\frac{1}{N}\sum_{j=1}^{N} z_j$,
$z=[z_1^T,\cdots,z_N^T]^T$, and
$e=[e_1^T,\cdots,e_N^T]^T$. Then, we get
$e=[(I_N-\frac{1}{N}\mathbf{1}\mathbf{1}^T)\otimes I_{2n}]z.$
It is easy
to see that $0$ is a simple eigenvalue of
$I_N-\frac{1}{N}\mathbf{1}\mathbf{1}^T$ with $\mathbf{1}$ as a
corresponding eigenvector, and 1 is the other eigenvalue with
multiplicity $N-1$. Then, it follows that $e=0$ if and only if
$z_1=\cdots=z_N$.
Therefore, the consensus problem of agents
\dref{1c} under the protocol \dref{clc3} is solved if
$e$ converges to
zero. It is not difficult to obtain that $e_i$
and $c_{ij}$ satisfy
\begin{equation}\label{netce2}
\begin{aligned}
\dot{e}_i &= \mathcal {M}e_i +\sum_{j=1}^{N}c_{ij}a_{ij}\mathcal {H}(e_i-e_j),\\
\dot{c}_{ij}&
=\kappa_{ij}a_{ij}(e_i-e_j)^T\mathcal {R}(e_i-e_j),\quad i=1,\cdots,N,
\end{aligned}
\end{equation}
where
$$\begin{aligned}
\mathcal {M} &=\begin{bmatrix}A & BF\\0 &
A+BF\end{bmatrix},\quad \mathcal {H}=\begin{bmatrix}0 & 0\\
-LC & LC\end{bmatrix},\\
\mathcal {R} &=(I_2\otimes C^T)\Gamma(I_2\otimes C).
\end{aligned}$$

The following theorem presents a sufficient condition for solving
the consensus problem.

{\bf Theorem 1}. Assume that the communication graph
$\mathcal {G}$ is undirected and connected. Then, the $N$ agents in
\dref{1c} reach consensus under the adaptive protocol \dref{clc3} with
$F$ satisfying that $A+BF$ is Hurwitz, $\Gamma=\begin{bmatrix}I_q & -I_q\\-I_q &I_q\end{bmatrix}$,
and $L=-Q^{-1}C^T$, where
$Q>0$ is a solution to
the following linear matrix inequality (LMI):
\begin{equation}\label{alg2}
A^TQ+QA-2C^TC<0.
\end{equation}
Moreover, the protocol states $v_i$, $i=1,\cdots,N$,
converge to zero and
each coupling weight $c_{ij}$ converges to some finite
steady-state value.

{\bf Proof.} Consider the Lyapunov function candidate
\begin{equation}\label{lyat21}
V_1(t)=\sum_{i=1}^{N}e_i^T\mathcal {Q}e_i+
\sum_{i=1}^{N}\sum_{j=1,j\neq i}^{N}\frac{(c_{ij}-\alpha)^2}{2\kappa_{ij}},
\end{equation}
where $\mathcal {Q}=\begin{bmatrix}\varsigma P+Q & -Q\\
-Q & Q\end{bmatrix}$, $P>0$ satisfies $P(A+BF)+(A+BF)^TP<0$, and
$\alpha$ and $\varsigma$ are positive constants
to be determined later. It is easy to verify that $\mathcal {Q}>0$.

The time derivative of $V_1(t)$ along the trajectory of \dref{netce2} can be obtained as
\begin{equation}\label{lyaet22}
\begin{aligned}
\dot{V}_1
&=2\sum_{i=1}^{N}e_i^T\mathcal {Q}\dot{e}_i+\sum_{i=1}^{N}\sum_{j=1,j\neq i}^{N}\frac{c_{ij}-\alpha}{\kappa_{ij}}\dot{c}_{ij}\\
&=2\sum_{i=1}^{N}e_i^T\mathcal {Q}[\mathcal {M}e_i +\sum_{j=1}^{N}c_{ij}a_{ij}\mathcal {H}(e_i-e_j)]\\
   &\quad +\sum_{i=1}^{N}\sum_{j=1,j\neq i}^{N}(c_{ij}-\alpha)a_{ij}(e_i-e_j)^T\mathcal {R}(e_i-e_j).
\end{aligned}
\end{equation}
Let $\tilde{e}_i=Te_i$, $i=1,\cdots,N$, with $T=\begin{bmatrix}I_q & 0\\
-I_q & I_q\end{bmatrix}$.
Then, \dref{lyaet22} can be rewritten as
\begin{equation}\label{lyaet23}
\begin{aligned}
\dot{V}_1
&=2\sum_{i=1}^{N}\tilde{e}_i^T\widetilde{\mathcal {Q}}[\widetilde{\mathcal {M}}\tilde{e}_i+\sum_{i=1}^{N}\sum_{j=1}^Nc_{ij}a_{ij}\widetilde{\mathcal {H}}(\tilde{e}_i-\tilde{e}_j)]\\
    &\quad+\sum_{i=1}^{N}\sum_{j=1,j\neq i}^{N}(c_{ij}-\alpha)a_{ij}(\tilde{e}_i-\tilde{e}_j)^T \widetilde{\mathcal {R}}(\tilde{e}_i-\tilde{e}_j),
\end{aligned}
\end{equation}
where
$$
\begin{aligned}
\widetilde{\mathcal {Q}} &\triangleq T^{-T}\mathcal {Q}T^{-1}=\begin{bmatrix}\varsigma P & 0\\ 0 & Q\end{bmatrix},\quad
\widetilde{\mathcal {M}}\triangleq T\mathcal {M}T^{-1}=\begin{bmatrix}A+BF & BF\\ 0 & A\end{bmatrix},\\
\widetilde{\mathcal {H}} &\triangleq T\mathcal {H}T^{-1}=\begin{bmatrix}0 & 0\\ 0 & LC\end{bmatrix},\quad
\widetilde{\mathcal {R}} \triangleq T^{-T}\mathcal {R} T^{-1}=  \begin{bmatrix}0 & 0\\ 0 & C^TC\end{bmatrix}.
\end{aligned}
$$
By noting that $L=-Q^{-1}C^T$, we can see that
\begin{equation}\label{lnot}
\widetilde{\mathcal {Q}}\widetilde{\mathcal {H}}=\widetilde{\mathcal {R}}.
\end{equation}

Because $\kappa_{ij}=\kappa_{ji}$, $c_{ij}(0)=c_{ji}(0)$, and $\Gamma$ is symmetric,
it follows from \dref{clc3} that
$c_{ij}(t)=c_{ji}(t)$, $\forall\,t\geq 0$. Therefore, we have
\begin{equation}\label{equa1}
\begin{aligned}
&\sum_{i=1}^{N}\sum_{j=1,j\neq i}^{N}(c_{ij}-\alpha)a_{ij}
(\tilde{e}_i-\tilde{e}_j)^T\widetilde{\mathcal {R}}(\tilde{e}_i-\tilde{e}_j)\\
&\quad\qquad=2\sum_{i=1}^{N}\sum_{j=1}^{N}(c_{ij}-\alpha)a_{ij}\tilde{e}_i^T\widetilde{\mathcal {R}}
(\tilde{e}_i-\tilde{e}_j).
\end{aligned}
\end{equation}
Let $\tilde{e}=[\tilde{e}_1^T,\cdots,\tilde{e}_N^T]^T$.
Using \dref{equa1} and \dref{lnot}, it follows from \dref{lyaet23} that
\begin{equation}\label{lyaet231}
\begin{aligned}
\dot{V}_1
&=2\sum_{i=1}^{N}\tilde{e}_i^T\widetilde{\mathcal {Q}}\widetilde{\mathcal {M}}\tilde{e}_i-2\sum_{i=1}^{N}\sum_{j=1}^{N}\alpha
a_{ij}\tilde{e}_i^T\widetilde{\mathcal {R}}(\tilde{e}_i-\tilde{e}_j)\\
&=\tilde{e}^T[I_N\otimes(\widetilde{\mathcal {Q}}\widetilde{\mathcal {M}}
+\widetilde{\mathcal {M}}^T\widetilde{\mathcal {Q}})
-2\alpha\mathcal {L}\otimes \widetilde{\mathcal {R}}]\tilde{e},
\end{aligned}
\end{equation}
where $\mathcal {L}$ is the Laplacian matrix associated with $\mathcal {G}$.

By the definitions of $e$ and $\tilde{e}$, it is easy to see that
$({\bf 1}^T\otimes I)\tilde{e}=({\bf 1}^T\otimes T)e=0$. Because $\mathcal {G}$ is connected,
it then follows from
Lemma 1 that
\begin{equation}\label{equa2}
\begin{aligned}
\tilde{e}^T(\mathcal {L}\otimes I)\tilde{e}\geq\lambda_2(\mathcal {L})\tilde{e}^T\tilde{e},
\end{aligned}
\end{equation}
where $\lambda_2(\mathcal {L})$ is the smallest nonzero eigenvalue of
$\mathcal {L}$. Therefore, we can get from \dref{lyaet231} that
\begin{equation}\label{lyaet24}
\begin{aligned}
\dot{V}_1
& \leq \tilde{e}^T[I_N\otimes(\widetilde{\mathcal {Q}}\widetilde{\mathcal {M}}
+\widetilde{\mathcal {M}}^T\widetilde{\mathcal {Q}}
-2\alpha\lambda_2(\mathcal {L})\widetilde{\mathcal {R}})]\tilde{e}.
\end{aligned}
\end{equation}

Note that
\begin{equation}\label{lyaet25}
\begin{aligned}
&\widetilde{\mathcal {Q}}\widetilde{\mathcal {M}}
+\widetilde{\mathcal {M}}^T\widetilde{\mathcal {Q}}
-2\alpha\lambda_2(\mathcal {L})\widetilde{\mathcal {R}}\\
&=\begin{bmatrix}
\varsigma(P(A+BF)+(A+BF)^TP) & \varsigma PBF\\
\varsigma F^TB^TP & QA+A^TQ-2\alpha\lambda_2(\mathcal {L})C^TC\end{bmatrix}.
\end{aligned}
\end{equation}
By choosing $\alpha$ sufficiently large such that
$\alpha\lambda_2(\mathcal {L})\geq1$, it follows from \dref{alg2}
that $QA+A^TQ-2\alpha\lambda_2(\mathcal {L})C^TC<0$.
Then, choosing $\varsigma>0$ sufficiently small,
and in virtue of Schur Complement Lemma \cite{boyd1994linear},
we can obtain from \dref{lyaet25} that
$$\widetilde{\mathcal {Q}}\widetilde{\mathcal {M}}
+\widetilde{\mathcal {M}}^T\widetilde{\mathcal {Q}}
-2\alpha\lambda_2(\mathcal {L})\widetilde{\mathcal {R}}<0.$$
Therefore, $\dot{V}_1\leq 0$.

Since $\dot{V}_1\leq 0$, $V_1(t)$ is bounded, implying that each $c_{ij}$
is also bounded.
By noting that $\mathcal {R}$ is
positive semi-definite, we can see from \dref{netce2}
that $c_{ij}$ is monotonically increasing.
Then, it follows that each coupling weight $c_{ij}$ converges to some finite value.
Note that $\dot{V}_1\equiv0$
implies that $\tilde{e}=0$ and $e=0$.
Hence, by LaSalle's Invariance principle
\cite{krstic1995nonlinear}, it follows that $e(t)\rightarrow 0$, as
$t\rightarrow \infty$. That is, the consensus problem is solved.
By \dref{clc3} and noting
the fact that $A+BF$ is Hurwitz,
it is easy to see that
the protocol states $v_i$, $i=1,\cdots,N$, converge to zero.
\hfill $\blacksquare$

{\bf Remark 1}.
As shown in \cite{li2010consensus}, a
necessary and sufficient condition for the existence of a $Q>0$ to
the LMI \dref{alg2} is that $(A,C)$ is detectable. Therefore, a
sufficient condition for the existence of a protocol \dref{clc3}
satisfying Theorem 1 is that $(A,B,C)$ is stabilizable and detectable.
It is worth noting that a favorable feature of \dref{clc3}
is that
its gain matrices $F$, $L$ and $\Gamma$
can be independently designed.

\subsection{Consensus Under Adaptive Protocol \dref{clc4}}

This section considers the consensus problem of the agents in \dref{1c} under
the adaptive protocol \dref{clc4}.
Let $\tilde{z}_i=[x_i^T,\tilde{v}_i^T]^T$,
$\zeta_i=\tilde{z}_i-\frac{1}{N}\sum_{j=1}^{N} \tilde{z}_j$, and $\zeta=[\zeta_1^T,\cdots,\zeta_N^T]^T$.
As shown in
the last subsection, the consensus problem of agents
\dref{1c} under the protocol \dref{clc4} is solved if
$\zeta$ converges to
zero. We can obtain that $\zeta_i$ and $d_i$ satisfy
the following dynamics:
\begin{equation}\label{netna}
\begin{aligned}
\dot{\zeta}_i &= \mathcal {M}\zeta_i +d_{i}\sum_{j=1}^{N}a_{ij}\mathcal {H}(\zeta_i-\zeta_j)
-\frac{1}{N}\sum_{k=1}^{N}d_k\sum_{j=1}^{N}a_{kj}\mathcal {H}(\zeta_k-\zeta_j),\\
\dot{d}_{i}&
=\tau_{i}[\sum_{j=1}^Na_{ij}(\zeta_i-\zeta_j)^T]\mathcal {R}[\sum_{j=1}^Na_{ij}(\zeta_i-\zeta_j)],\quad i=1,\cdots,N,
\end{aligned}
\end{equation}
where $\mathcal {M}$, $\mathcal {H}$, $\mathcal {R}$
are defined in \dref{netce2}. Let
$D(t)=\rm{diag}(d_1(t),\cdots,d_N(t))$.
Then, the first equation in \dref{netna} can be rewritten into a compact form as
\begin{equation}\label{netna2}
\begin{aligned}
\dot{\zeta} &= [I_N\otimes \mathcal {M}+((I_N-\frac{1}{N}{\bf 1}{\bf 1}^T)D\mathcal {L})\otimes\mathcal {H}]\zeta,
\end{aligned}
\end{equation}
where $\mathcal {L}$ is the Laplacian matrix associated with $\mathcal {G}$.

{\bf Theorem 2}. Assume that the communication graph
$\mathcal {G}$ is undirected and connected. Then, the $N$ agents in
\dref{1c} reach consensus under the protocol \dref{clc4} with
$F$, $L$, and $\Gamma$ given as in Theorem 1.
Moreover, the protocol states $\tilde{v}_i$, $i=1,\cdots,N$,
converge to zero and
each coupling weight $\tilde{d}_{i}$ converges to some finite
steady-state value.

{\bf Proof.} Consider the Lyapunov function candidate
\begin{equation}\label{lyan1}
V_2=\zeta^T(\mathcal {L}\otimes \mathcal {Q})\zeta+
\sum_{i=1}^{N}\frac{(d_{i}-\beta)^2}{2\tau_{i}},
\end{equation}
where $\mathcal {Q}$ is defined in \dref{lyat21}, and
$\beta$ is a positive constant
to be determined later. For a connected graph $\mathcal {G}$,
it follows from Lemma 1 and the definition of $\zeta$ that
$\zeta^T(\mathcal {L}\otimes \mathcal {Q})\zeta\geq
\lambda_2(\mathcal {L})\zeta^T(I_N\otimes \mathcal {Q})\zeta$.
Therefore, it is easy to see that $\Omega_c=\{\zeta,d_i|V_2\leq c\}$ is compact for
any positive $c$.

Following similar steps to those in the proof of Theorem 1,
we can obtain the time derivative of $V_2$ along the trajectory of \dref{netna2} as
\begin{equation}\label{lyan2}
\begin{aligned}
\dot{V}_2
&=2\zeta^T(\mathcal {L}\otimes \mathcal {Q})\zeta+\sum_{i=1}^{N}\frac{d_{i}-\beta}{\tau_{i}}\dot{d}_{i}\\
&=2\zeta^T[\mathcal {L}\otimes \mathcal {Q}\mathcal {M}+(\mathcal {L}D\mathcal {L}-\mathcal {L}
{\bf 1}{\bf 1}^TD\mathcal {L})\otimes\mathcal {Q}\mathcal {H}]\zeta+\sum_{i=1}^{N}\frac{d_{i}-\beta}{\tau_{i}}\dot{d}_{i}\\
&=2\tilde{\zeta}^T[\mathcal {L}\otimes \widetilde{\mathcal
{Q}}\widetilde{\mathcal {M}}+(\mathcal {L}D\mathcal
{L})\otimes\widetilde{\mathcal
{R}}]\tilde{\zeta}+\sum_{i=1}^{N}\frac{d_{i}-\beta}{\tau_{i}}\dot{d}_{i}.
\end{aligned}
\end{equation}
where $\tilde{\zeta}\triangleq[\tilde{\zeta}^T_1,\cdots,\tilde{\zeta}^T_N]^T=(I_N\otimes T)\zeta$
with $T=\begin{bmatrix}I_q & 0\\
-I_q & I_q\end{bmatrix}$, and
$\widetilde{\mathcal {Q}}$, $\widetilde{\mathcal {M}}$, $\widetilde{\mathcal {R}}$ are the same as in \dref{lyaet23}.
Observe that
\begin{equation}\label{lyan3}
\begin{aligned}
\tilde{\zeta}^T[(\mathcal {L}D\mathcal {L})\otimes\widetilde{\mathcal {R}}]\tilde{\zeta} =\sum_{i=1}^{N}d_{i}[\sum_{j=1}^{N}a_{ij}(\tilde{\zeta}_i-\tilde{\zeta}_j)^T]\widetilde{\mathcal {R}}
[\sum_{j=1}^{N}a_{ij}(\tilde{\zeta}_i-\tilde{\zeta}_j)].
\end{aligned}
\end{equation}
Moreover, the second equation in \dref{netna} can be rewritten as
\begin{equation}\label{lyan4}
\begin{aligned}
\dot{d}_{i}
=\tau_{i}[\sum_{j=1}^Na_{ij}(\tilde{\zeta}_i-\tilde{\zeta}_j)^T]\widetilde{\mathcal {R}}
[\sum_{j=1}^Na_{ij}(\tilde{\zeta}_i-\tilde{\zeta}_j)].
\end{aligned}
\end{equation}
Substituting \dref{lyan3} and \dref{lyan4} into \dref{lyan2} yields
\begin{equation}\label{lyan5}
\begin{aligned}
\dot{V}_2
&=2\tilde{\zeta}^T(\mathcal {L}\otimes \widetilde{\mathcal {Q}}\widetilde{\mathcal {M}})\tilde{\zeta}
-2\beta\sum_{i=1}^N[\sum_{j=1}^Na_{ij}(\tilde{\zeta}_i-\tilde{\zeta}_j)^T]\widetilde{\mathcal {R}}
[\sum_{j=1}^Na_{ij}(\tilde{\zeta}_i-\tilde{\zeta}_j)]\\
&=\tilde{\zeta}^T[\mathcal {L}\otimes (\widetilde{\mathcal
{Q}}\widetilde{\mathcal {M}} +\widetilde{\mathcal
{M}}^T\widetilde{\mathcal {Q}}^T)-2\beta\mathcal
{L}^2\otimes\widetilde{\mathcal {R}}]\tilde{\zeta}.
\end{aligned}
\end{equation}

Because $\mathcal {G}$ is connected, it follows from Lemma 1 that
zero is a simple eigenvalue of
$\mathcal {L}$ and all the other eigenvalues are positive.
Let $U=\left[\begin{smallmatrix}
\mathbf{1} & Y_1
\end{smallmatrix}\right]$ and $U^T=\left[\begin{smallmatrix}
\frac{\mathbf{1}^T}{N} \\ Y_2
\end{smallmatrix}\right]$, with $Y_1\in\mathbf{R}^{N\times(N-1)}$, $Y_2\in\mathbf{R}^{(N-1)\times N}$,
be such unitary matrices that
$U^{T}\mathcal {L}U=\Lambda\triangleq{\rm{diag}}(0,\lambda_2,\cdots,\lambda_N)$,
where $\lambda_2\leq\cdots\leq\lambda_N$ are the nonzero eigenvalues of $\mathcal {L}$.
Let $\bar{\zeta}\triangleq[\bar{\zeta}_1^T,\cdots,\bar{\zeta}_N^T]^T=(U^T\otimes I)\tilde{\zeta}$.
By the definitions of $\zeta$ and $\tilde{\zeta}$, it is easy to see that
$
\bar{\zeta}_1=(\mathbf{1}^T\otimes T)\zeta=0.
$
Then, it follows from \dref{lyan5} that
\begin{equation}\label{lyan6}
\begin{aligned}
\dot{V}_2 &=\bar{\zeta}^T[\Lambda\otimes (\widetilde{\mathcal {Q}}\widetilde{\mathcal {M}}
+\widetilde{\mathcal {M}}^T\widetilde{\mathcal {Q}}^T)-2\beta\Lambda^2\otimes\widetilde{\mathcal {R}}]\bar{\zeta}\\
&=\sum_{i=2}^{N}\lambda_i\bar{\zeta}_i^T(\widetilde{\mathcal {Q}}\widetilde{\mathcal {M}}
+\widetilde{\mathcal {M}}^T\widetilde{\mathcal {Q}}^T-2\beta\lambda_i\widetilde{\mathcal {R}})\bar{\zeta}_i.
\end{aligned}
\end{equation}

As shown in the proof of Theorem 1, by choosing $\beta$ sufficiently large such that
$\beta\lambda_2(\mathcal {L})\geq1$ and $\varsigma>0$ sufficiently small,
we can obtain from \dref{lyan6} that $\dot{V}_2\leq 0$. Note that
$\dot{V}_2\equiv 0$ that $\bar{\zeta}_i=0$, $i=2,\cdots,N$,
which, together with $\bar{\zeta}_1=0$, further implies that $\zeta=0$. Therefore,
it follows from Lasalle's Invariance principle that $\zeta\rightarrow 0$, as $t\rightarrow\infty$.
The convergence of $d_i$ and $\tilde{v}_i$, $i=1,\cdots,N$, 
can be shown by following similar steps in the proof of Theorem 1,
which is omitted here for brevity.
\hfill $\blacksquare$

{\bf Remark 2}. Different from the previous adaptive schemes in
\cite{li2011adaptive,de2008synchronization,yu2011distributed}, which
are based on the relative state information, the proposed adaptive
protocols \dref{clc3} and \dref{clc4} rely on the relative outputs
of neighboring agents. Contrary to the protocols in
\cite{li2010consensus,li2011consensusSCL,seo2009consensus,zhang2011optimal,ma2010necessary},
the adaptive protocols \dref{clc3} and \dref{clc4} can be computed
and implemented by each agent in a fully distributed fashion without
using any global information. It is worth mentioning that the
consensus protocols in
\cite{tuna2009conditions,scardovi2009synchronization} do not need
any global information either. However, in \cite{tuna2009conditions}
the protocol is based on the relative states of neighboring agents
and the agent dynamics are restricted to be neutrally stable. In
\cite{scardovi2009synchronization}, the eigenvalues of the state
matrix of each agent are assumed to lie in the closed left-half
plane. Furthermore, the dimension of the protocol in
\cite{scardovi2009synchronization} is higher than that of the
adaptive protocol \dref{clc4}.

{\bf Remark 3}. Some comparisons between the adaptive consensus protocols \dref{clc3} and
\dref{clc4} are now briefly discussed.
The dimension of the adaptive protocol \dref{clc3}
is proportional to the number of edges in the communication graph.
Since the number of edges is usually larger than the number of nodes
in a connected graph, the dimension of the protocol \dref{clc3}
are generally higher than that of \dref{clc4}.
On the other hand, the adaptive protocol \dref{clc3}
are applicable to the case with switching communication graphs,
which will be shown in Section 5.

\section{Consensus with Leader-Follower Communication Graphs}

The section extends to consider the case where
the $N$ agents in \dref{1c} maintain a leader-follower communication graph $\mathcal {G}$.
Without loss of generality, assume that the agent indexed by
1 is the leader whose control input
$u_1=0$ and the agents indexed by $2,\cdots,N$, are followers.
The leader does not receive any information from the followers,
i.e., it has no neighbor, while
each follower can obtain the relative outputs with respect
to its neighbors.

In the sequel, the following assumption is needed.

{\bf Assumption 1}.
The subgraph associated with the followers is undirected and
the graph $\mathcal {G}$ contains a directed spanning tree
with the leader as the root.

Denote by $\mathcal {L}$ the Laplacian matrix associated with
$\mathcal {G}$. Because the leader has no neighbors, $\mathcal
{L}$ can be partitioned as
\begin{equation}\label{lapc}
\mathcal {L}=\begin{bmatrix} 0 & 0_{1\times (N-1)} \\
\mathcal {L}_2 & \mathcal {L}_1\end{bmatrix},
\end{equation}
where $\mathcal
{L}_2\in\mathbf{R}^{(N-1)\times 1}$ and $\mathcal {L}_1\in\mathbf{R}^{(N-1)\times (N-1)}$
is symmetric.

It is said that the leader-follower consensus problem is solved if the states of
the followers converge to the state of the leader, i.e.,
$\lim_{t\rightarrow
\infty}\|x_i(t)- x_1(t)\|=0$, $ \forall\,i=2,\cdots,N$.

Based on the relative output information of neighboring agents, the
first adaptive consensus protocol with
a dynamic coupling weight for each edge is proposed for the followers as follows:
\begin{equation}\label{clcf}
\begin{aligned}
\dot{\bar{v}}_i &=(A+BF)\bar{v}_i+L\sum_{j=1}^{N}d_{ij}a_{ij}
\left[C(\bar{v}_i-\bar{v}_j)-(y_i-y_j)\right],\\
\dot{d}_{ij}&
=\epsilon_{ij}a_{ij}\begin{bmatrix}y_i-y_j \\ C(\bar{v}_i-\bar{v}_j)\end{bmatrix}^T\Gamma
\begin{bmatrix}y_i-y_j \\ C(\bar{v}_i-\bar{v}_j)\end{bmatrix},\\
u_i &
=F\bar{v}_i,\quad i=2,\cdots,N,
\end{aligned}
\end{equation}
where $\bar{v}_i\in\mathbf{R}^{n}$ is the protocol state, $i=2,\cdots,N$,
$\bar{v}_1\in\mathbf{R}^{n}$ is generated by
$\dot{\bar{v}}_1=(A+BF)\bar{v}_1$, $a_{ij}$ is the $(i,j)$-th entry of
the adjacency matrix $\mathcal {A}$ of
$\mathcal {G}$, $d_{ij}$
is the coupling weight associated with
the edge $(j,i)$
with $d_{ij}(0)=d_{ji}(0)$ for $i,j=2,\cdots,N$,
$\epsilon_{ij}=\epsilon_{ji}$ are positive constants,
$L\in\mathbf{R}^{q\times n}$, $F\in\mathbf{R}^{p\times n}$,
and $\Gamma\in\mathbf{R}^{2q\times 2q}$.

{\bf Theorem 3}. For any graph
$\mathcal {G}$ satisfying Assumption 1, the $N$ agents in
\dref{1c} reach leader-follower
consensus under the protocol \dref{clcf} with
$F$,
$L$, and $\Gamma$ given as in Theorem 1. Meanwhile,
the protocol states $\bar{v}_i$, $i=2,\cdots,N$,
converge to zero and the coupling weights $d_{ij}$
converge to finite steady-state values.

{\bf Proof.}
Let $\xi_i=\begin{bmatrix}x_i-x_1\\\bar{v}_i-\bar{v}_1\end{bmatrix}$, $i=2,\cdots,N$.
Then, we can get from \dref{1c} and
\dref{clcf} that
\begin{equation}\label{netcef}
\begin{aligned}
\dot{\xi}_i &= \mathcal {M}\xi_i+\sum_{j=2}^{N}d_{ij}a_{ij}\mathcal {H}(\xi_i-\xi_j)+d_{i1}a_{i1}\mathcal {H}\xi_i,\\
\dot{d}_{i1}& =\epsilon_{i1}a_{i1}\xi_i^T\mathcal {R} \xi_i,\\
\dot{d}_{ij}& =\epsilon_{ij}a_{ij}(\xi_i-\xi_j)^T\mathcal {R}(\xi_i-\xi_j),
~i=2,\cdots,N,
\end{aligned}
\end{equation}
where $\mathcal {M}$, $\mathcal {H}$ and $\mathcal {R}$ are the same as in \dref{netce2}.
Clearly, the leader-follower consensus problem
of \dref{1c} is solved by \dref{clcf}
if the states $\xi_i$ of \dref{netcef}
converge to zero.

Consider the Lyapunov function candidate
\begin{equation}\label{lyaf1}
\begin{aligned}
V_3 =\sum_{i=2}^{N}\xi_i^T\mathcal {Q}\xi_i+ \sum_{i=2}^{N}\sum_{j=2,j\neq
i}^{N}\frac{(d_{ij}-\gamma)^2}{2\epsilon_{ij}}
+\sum_{i=2}^{N}\frac{(d_{i1}-\gamma)^2}{\epsilon_{i1}},
\end{aligned}
\end{equation}
where $\gamma$ is a positive constant and $\mathcal {Q}$
is defined in \dref{lyat21}. Following
similar steps to those in the proof of Theorem 1, the time derivative of $V_2$
along the trajectory of \dref{netcef} can be obtained as
\begin{equation}\label{lyaf2}
\begin{aligned}
\dot{V}_3
&=2\sum_{i=2}^{N}\xi_i^T\mathcal {Q}\dot{\xi}_i+\sum_{i=2}^{N}\sum_{j=2,j\neq i}^{N}\frac{d_{ij}-\beta}{\epsilon_{ij}}\dot{d}_{ij}
    +\sum_{i=2}^{N}\frac{2(d_{i1}-\gamma)}{\epsilon_{i1}}\dot{d}_{i1}\\
&=2\sum_{i=2}^{N}\tilde{\xi}_i^T\widetilde{\mathcal {Q}}[\widetilde{\mathcal {M}}\tilde{\xi}_i+\sum_{j=2}^{N}d_{ij}a_{ij}\widetilde{\mathcal {H}}(\tilde{\xi}_i-\tilde{\xi}_j)+d_{i1}a_{i1}\widetilde{\mathcal {H}}\tilde{\xi}_i]\\
    &\quad+\sum_{i=2}^{N}\sum_{j=2,j\neq i}^{N}(d_{ij}-\gamma)a_{ij}(\tilde{\xi}_i-\tilde{\xi}_j)^T\widetilde{\mathcal {R}}(\tilde{\xi}_i-\tilde{\xi}_j)+2\sum_{i=2}^{N}(d_{i1}-\gamma)a_{i1}\tilde{\xi}_i^T\widetilde{\mathcal {R}} \tilde{\xi}_i,
\end{aligned}
\end{equation}
where $\tilde{\xi}_i=T\xi_i$ with $T=\begin{bmatrix}I_q & 0\\
-I_q & I_q\end{bmatrix}$, and $\widetilde{\mathcal {Q}}$,
$\widetilde{\mathcal {M}}$, $\widetilde{\mathcal {H}}$,
$\widetilde{\mathcal {R}}$ are defined in \dref{lyaet23}.

Since the subgraph associated with the $N-1$ followers is undirected,
we have
$$
\begin{aligned}
&\sum_{i=2}^{N}\sum_{j=2,j\neq i}^{N}(d_{ij}-\gamma)a_{ij}
(\tilde{\xi}_i-\tilde{\xi}_j)^T\widetilde{\mathcal {R}}(\tilde{\xi}_i-\tilde{\xi}_j)\\
&\quad\qquad=2\sum_{i=2}^{N}\sum_{j=2}^{N}(d_{ij}-\gamma)a_{ij}\tilde{\xi}_i^T\widetilde{\mathcal {R}}
(\tilde{\xi}_i-\tilde{\xi}_j).
\end{aligned}
$$
Letting $\tilde{\xi}=[\tilde{\xi}_2^T,\cdots,\tilde{\xi}_N^T]^T$,
it then follows from \dref{lyaf2} that
\begin{equation}\label{lyaf3}
\begin{aligned}
\dot{V}_3
&=2\sum_{i=2}^{N}\tilde{\xi}_i^T\widetilde{\mathcal {Q}}\widetilde{\mathcal {M}}\tilde{\xi}_i-2\gamma\sum_{i=2}^{N}\sum_{j=2}^{N}a_{ij}\tilde{\xi}_i^T\widetilde{\mathcal {R}}(\tilde{\xi}_i-\tilde{\xi}_j)
-2\gamma\sum_{i=2}^{N}a_{i1}\tilde{\xi}_i^T\widetilde{\mathcal {R}} \tilde{\xi}_i\\
&=\tilde{\xi}^T[I_{N-1}\otimes(\widetilde{\mathcal {Q}}\widetilde{\mathcal {M}}
+\widetilde{\mathcal {M}}^T\widetilde{\mathcal {Q}})
-2\gamma\mathcal {L}_1\otimes \widetilde{\mathcal {R}}]\tilde{\xi},
\end{aligned}
\end{equation}
where $\mathcal {L}_1$ is defined in \dref{lapc}.

For any graph $\mathcal {G}$ satisfying Assumption 1,
it follows from Lemma 1 and \dref{lapc} that
$\mathcal {L}_1$ is positive definite. Thus,
$$\tilde{\xi}^T(\mathcal {L}_1\otimes \widetilde{\mathcal {R}})\tilde{\xi}\geq
\lambda_2(\mathcal {L})\tilde{\xi}^T(I_{N-1}\otimes \widetilde{\mathcal {R}})\tilde{\xi},$$
where $\lambda_2(\mathcal {L})$ is the smallest eigenvalue of $\mathcal {L}_1$.
Then, we have
\begin{equation}\label{lyaf4}
\begin{aligned}
\dot{V}_3
\leq\tilde{\xi}^T[I_{N-1}\otimes(\widetilde{\mathcal {Q}}\widetilde{\mathcal {M}}
+\widetilde{\mathcal {M}}^T\widetilde{\mathcal {Q}}
-2\gamma\lambda_2(\mathcal {L})\widetilde{\mathcal {R}})]\tilde{\xi}.
\end{aligned}
\end{equation}
The rest of the proof is similar to that of Theorem 1,
which is omitted for brevity.
\hfill $\blacksquare$

Corresponding to \dref{clc4}, the second adaptive consensus protocol
with a time-varying coupling weight for each follower is proposed as
\begin{equation}\label{clcf2}
\begin{aligned}
\dot{\hat{v}}_i &=(A+BF)\hat{v}_i+\hat{d}_iL\sum_{j=1}^{N}a_{ij}
\left[C(\hat{v}_i-\hat{v}_j)-(y_i-y_j)\right],\\
\dot{\hat{d}}_{i}&
=\epsilon_{i}\left(\sum_{j=1}^Na_{ij}\begin{bmatrix}y_i-y_j \\
C(\hat{v}_i-\hat{v}_j)\end{bmatrix}^T\right)\Gamma
\left(\sum_{j=1}^Na_{ij}\begin{bmatrix}y_i-y_j \\ C(\hat{v}_i-\hat{v}_j)\end{bmatrix}\right),\\
u_i &
=F\hat{v}_i,\quad i=2,\cdots,N,
\end{aligned}
\end{equation}
where $\hat{v}_i\in\mathbf{R}^{n}$ is the protocol state, $i=2,\cdots,N$,
$\hat{v}_1\in\mathbf{R}^{n}$ is generated by
$\dot{\hat{v}}_1=(A+BF)\hat{v}_1$, $\hat{d}_{i}$ denotes
the coupling weight associated with follower $i$, and
$\epsilon_{i}$ are positive constants.

{\bf Theorem 4}. For any graph
$\mathcal {G}$ satisfying Assumption 1, the $N$ agents in
\dref{1c} reach leader-follower
consensus under the protocol \dref{clcf2} with
with $F$,
$L$, and $\Gamma$ given as in Theorem 1.
Meanwhile,
the protocol states $\hat{v}_i$, $i=2,\cdots,N$,
converge to zero and each coupling weight $\hat{d}_{i}$
converges to some finite steady-state value.

{\bf Proof.}
Let $\varrho_i=\begin{bmatrix}x_i-x_1\\\hat{v}_i-\hat{v}_1\end{bmatrix}$, $i=2,\cdots,N$.
Clearly, the leader-follower consensus problem
of \dref{1c} under \dref{clcf2} is solved
if $\varrho_i$, $i=2,\cdots,N$,
converge to zero. It is easy to see that $\varrho_i$
and $\hat{d}_i$ satisfy
\begin{equation}\label{netcn2}
\begin{aligned}
\dot{\varrho}_i &= \mathcal {M}\varrho_i+\hat{d}_{i}\mathcal {H}[\sum_{j=2}^{N}a_{ij}(\varrho_i-\varrho_j)+a_{i1}\varrho_i],\\
\dot{\hat{d}}_{i}& =\epsilon_{i}[\sum_{j=2}^{N}a_{ij}(\varrho_i-\varrho_j)+a_{i1}\varrho_i]^T\mathcal {R} [\sum_{j=2}^{N}a_{ij}(\varrho_i-\varrho_j)+a_{i1}\varrho_i],
~i=2,\cdots,N.
\end{aligned}
\end{equation}

Consider the Lyapunov function candidate
\begin{equation}\label{lyafn1}
\begin{aligned}
V_4 =\sum_{i=2}^{N}\varrho_i^T\mathcal {Q}\varrho_i+ \sum_{i=2}^{N}\frac{(\hat{d}_{i}-\sigma)^2}{2\epsilon_{i}},
\end{aligned}
\end{equation}
where $\sigma$ is a positive constant and $\mathcal {Q}$
is defined in \dref{lyat21}.
The rest of the proof can be completed by following similar steps in proving Theorems 2 and 3.
\hfill $\blacksquare$


\section{Extensions to Switching Communication Graphs}

In the last sections, the communication graph is assumed to be fixed
throughout the whole process. However, the communication graph may
change with time in many practical situations due to various
reasons, such as communication constraints, link
variations, etc. In this section, the consensus problem under the
adaptive protocol \dref{clc3} with switching communication graphs
will be considered.
%

Denote by $\mathscr{G}_N$ the set of all possible undirected connected
graphs with $N$ nodes. Let
$\sigma(t):[0,\infty)\rightarrow \mathscr{P}$ be a piecewise constant switching signal
with switching times $t_0,t_1,\cdots$,
and $\mathscr{P}$ be the index set associated with the elements of $\mathscr{G}_N$, which is
clearly finite. The communication graph at time $t$ is denoted by $\mathcal {G}_{\sigma(t)}$.
Accordingly, \dref{clc3} becomes
\begin{equation}\label{clcsw}
\begin{aligned}
\dot{v}_i &=(A+BF)v_i+L\sum_{j=1}^Nc_{ij}a_{ij}(t)\left[C(v_i-v_j)-(y_i-y_j)\right],\\
\dot{c}_{ij}&
=\kappa_{ij}a_{ij}(t)\begin{bmatrix}y_i-y_j \\ C(v_i-v_j)\end{bmatrix}^T\Gamma
\begin{bmatrix}y_i-y_j \\ C(v_i-v_j)\end{bmatrix},\\
u_i &
=Fv_i,\quad i=1,\cdots,N,
\end{aligned}
\end{equation}
where $a_{ij}(t)$ is the $(i,j)$-th entry of
the adjacency matrix associated with $\mathcal {G}_{\sigma(t)}$
and the rest of the variables are the same as in \dref{clc3}.

{\bf Theorem 5}. For arbitrary switching communication graphs
$\mathcal {G}_{\sigma(t)}$
belonging
to $\mathscr{G}_N$, the $N$ agents in
\dref{1c} reach consensus under the protocol \dref{clcsw} with
$F$,
$L$ and $\Gamma$ given as in Theorem 1. Besides, the protocol states $v_i$,
$i=1,\cdots,N$,
converge to zero and the coupling weights
$c_{ij}$ converge to some finite values.

{\bf Proof.}
Let $e_i=x_i-\frac{1}{N}\sum_{j=1}^{N} x_j$, $i=1,\cdots,N$,
and $e=[e_1^T,\cdots,e_N^T]^T$.
By following similar steps to those in the proof of Theorem 1,
the consensus problem of the agents
\dref{1c} under the protocol \dref{clcsw} is solved if $e$
converges to zero. Clearly $e_i$ and $\tilde{c}_{ij}$ satisfy
\begin{equation}\label{netsw}
\begin{aligned}
\dot{e}_i &= \mathcal {M}e_i +\sum_{j=1}^{N}(\tilde{c}_{ij}+\delta)a_{ij}(t)\mathcal {H}(e_i-e_j),\\
\dot{\tilde{c}}_{ij}&
=\kappa_{ij}a_{ij}(t)(e_i-e_j)^T\mathcal {R}(e_i-e_j),\quad i=1,\cdots,N,
\end{aligned}
\end{equation}
where $c_{ij}=\tilde{c}_{ij}+\delta$, $\delta$ is a positive scalar,
and $\mathcal {M}$, $\mathcal {H}$ and $\mathcal {R}$ are defined as in \dref{netce2}.

Take a common Lyapunov function candidate as
\begin{equation}\label{lyasw1}
V_5(t)=\sum_{i=1}^{N}e_i^T\mathcal {Q}e_i+
\sum_{i=1}^{N}\sum_{j=1,j\neq i}^{N}\frac{\tilde{c}_{ij}^2}{2\kappa_{ij}},
\end{equation}
where $\mathcal {Q}$ is defined in \dref{lyat21}.
The time derivative of $V_5$ along the trajectory of \dref{netsw} can be obtained as
\begin{equation}\label{lyasw2}
\begin{aligned}
\dot{V}_5
&=2\sum_{i=1}^{N}e_i^T\mathcal {Q}\dot{e}_i+\sum_{i=1}^{N}\sum_{j=1,j\neq i}^{N}\frac{\tilde{c}_{ij}}{\kappa_{ij}}\dot{\tilde{c}}_{ij}\\
&=2\sum_{i=1}^{N}e_i^T\widetilde{\mathcal {Q}}[\widetilde{\mathcal {M}}e_i +\sum_{j=1}^{N}(\tilde{c}_{ij}+\delta)a_{ij}(t)\widetilde{\mathcal {H}}(e_i-e_j)]\\
   &\quad +\sum_{i=1}^{N}\sum_{j=1,j\neq i}^{N}c_{ij}a_{ij}(t)(e_i-e_j)^T\widetilde{\mathcal {R}}(e_i-e_j)\\
&=\tilde{e}^T[I_N\otimes(\widetilde{\mathcal {Q}}\widetilde{\mathcal {M}}
+\widetilde{\mathcal {M}}^T\widetilde{\mathcal {Q}})
-2\delta\mathcal {L}_{\sigma(t)}\otimes \widetilde{\mathcal {R}}]\tilde{e},
\end{aligned}
\end{equation}
where \dref{equa1} has been used to obtain the last equality,
$\mathcal {L}_{\sigma(t)}$ is the Laplacian matrix associated with $\mathcal {G}_{\sigma(t)}$,
and the rest of the variables are the same as in \dref{lyaet23}.

Since $\mathcal {G}_{\sigma(t)}$ is connected and
$({\bf 1}^T\otimes I)\tilde{e}=0$, it is easy to see that
\begin{equation}\label{equasw}
\begin{aligned}
\tilde{e}^T(\mathcal {L}_{\sigma(t)}\otimes I)\tilde{e}\geq\lambda_2^{\min}\tilde{e}^T\tilde{e},
\end{aligned}
\end{equation}
where $\lambda_2^{\min}\triangleq
\min_{\mathcal {G}_{\sigma(t)}\in\mathscr{G}_N}\{\lambda_2(\mathcal {L}_{\sigma(t)})\}$
denotes the minimum of the smallest nonzero eigenvalues of
$\mathcal {L}_{\sigma(t)}$ for all $\mathcal {G}_{\sigma(t)}\in\mathscr{G}_N$.
Therefore, we can get from \dref{lyasw2} that
\begin{equation}\label{lyasw3}
\begin{aligned}
\dot{V}_5
& \leq \tilde{e}^T[I_N\otimes(\widetilde{\mathcal {Q}}\widetilde{\mathcal {M}}
+\widetilde{\mathcal {M}}^T\widetilde{\mathcal {Q}}
-2\delta\lambda_2^{\min}\widetilde{\mathcal {R}})]\tilde{e}\\
&\triangleq W(\tilde{e}).
\end{aligned}
\end{equation}

As shown in the proof of Theorem 1,
by choosing $\delta$ sufficiently large such that
$\delta\lambda_2^{\min}\geq1$ and $\gamma>0$ sufficiently small,
we have
$\widetilde{\mathcal {Q}}\widetilde{\mathcal {M}}
+\widetilde{\mathcal {M}}^T\widetilde{\mathcal {Q}}
-2\delta\lambda_2^{\min}\widetilde{\mathcal {R}}<0.$
Therefore, $\dot{V}_5\leq 0$,
implying that $V_5$ is bounded. Consequently, $c_{ij}$, $i,j=1,\cdots,N$,
are bounded. By noting \dref{netsw},
each $\tilde{c}_{ij}$ is monotonically increasing. It then follows that
each $\tilde{c}_{ij}$ converges to some finite value. Thus,
the coupling weights $c_{ij}$ converge to finite steady-state values.
Note that $V_5$ is positive definite and radically unbounded.
By LaSalle-Yoshizawa theorem \cite{krstic1995nonlinear},
it follows that $\lim_{t\rightarrow \infty}W(\tilde{e})=0$, implying
that $\tilde{e}(t)\rightarrow 0$, as $t\rightarrow \infty$,
which further
implies that $e(t)\rightarrow 0$, as $t\rightarrow \infty$.
This completes the proof.
\hfill $\blacksquare$

{\bf Remark 4}. Theorem 5 shows that the adaptive consensus
protocol \dref{clc3} given by Theorem 1 is applicable
to arbitrary switching communication graphs which are connected at
any time instant. The case with switching
leader-follower graphs can be discussed similarly, thus
is omitted here for brevity. Because the Lyapunov function in
\dref{lyan1} for the adaptive protocol \dref{clc4}
is explicitly related with the communication graph,
it cannot be taken as a feasible common Lyapunov function.

\section{Simulation Examples}

In this section,
a simulation example is provided to validate the
effectiveness of the theoretical results.

\begin{figure}[htbp]
\centering
\subfigure[$\mathcal {G}_1$]{\includegraphics[width=0.3\linewidth]{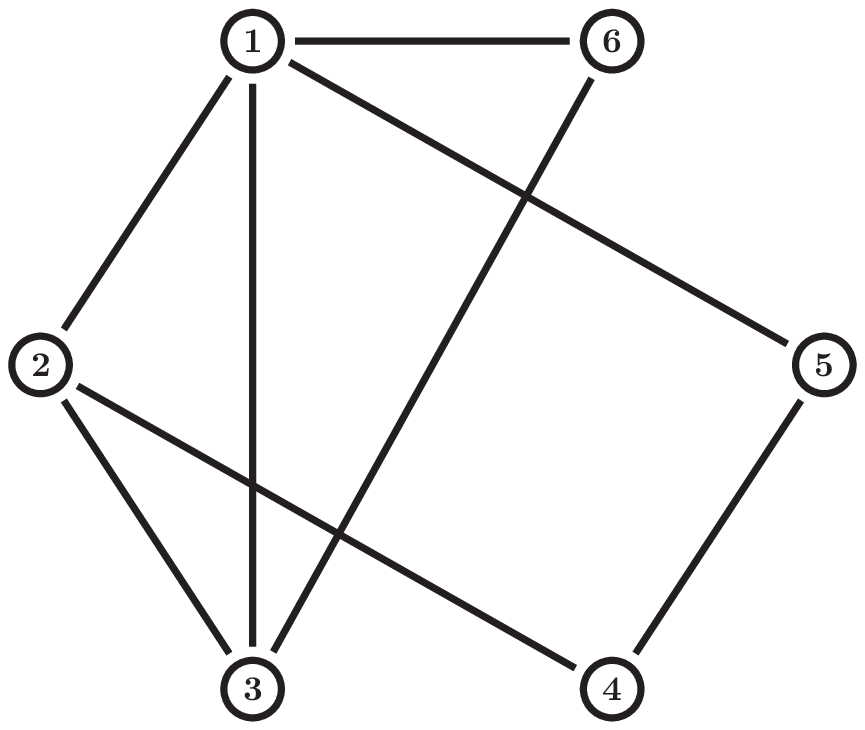}}\qquad\quad
\subfigure[$\mathcal {G}_2$]{
\includegraphics[width=0.3\linewidth]{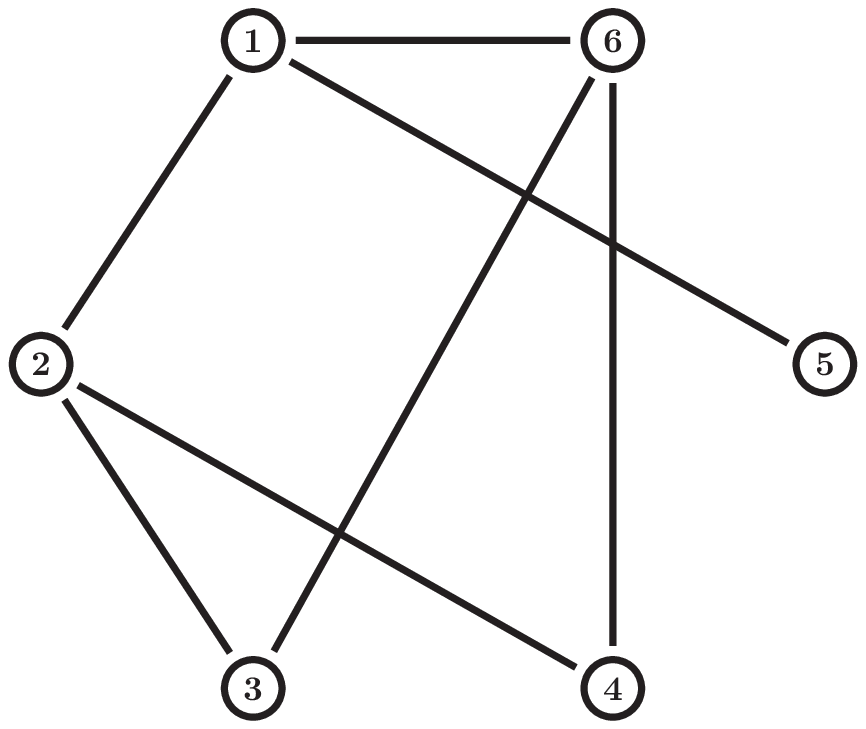}
}
\caption{Undirected communication graphs $\mathcal {G}_1$ and $\mathcal {G}_2$.}
\end{figure}

Consider a network of third-order integrators,
described by \dref{1c} with
$$\begin{aligned}
A=\begin{bmatrix}0 & 1& 0 \\ 0 & 0 & 1\\
0 &0 & 0 \end{bmatrix},\quad B
=\begin{bmatrix} 0 \\ 0 \\ 1\end{bmatrix},\quad
C=\begin{bmatrix} 1 & 0& 0\end{bmatrix}.
\end{aligned}$$

Choose $F=-\left[\begin{smallmatrix} 3 &  6.5 & 4.5\end{smallmatrix}\right]$ such that $A+BF$ is Hurwitz.
Solving the LMI \dref{alg2} by using the LMI toolbox of Matlab
gives the gain matrix $L$ in \dref{clcsw} and \dref{clc4} as
$L=-\left[\begin{smallmatrix} 2.5039 & 1.9056 & 0.9194\end{smallmatrix}\right]^T$.
To illustrate Theorem 5, let $\mathcal {G}_{\sigma(t)}$ switch
randomly every 0.1 second between $\mathcal {G}_1$ and $\mathcal {G}_2$ as shown in Figure 1.
Note that both $\mathcal {G}_1$ and $\mathcal {G}_2$ are connected.
Let $\kappa_{ij}=1$, $i,j=1,\cdots,6$, in \dref{clcsw}, and
$c_{ij}(0)=c_{ji}(0)$ be randomly chosen.
The consensus errors $x_i-x_1$, $i=2,\cdots,5$, of the third-order integrators
under the protocol \dref{clc3} with $F$, $L$ as above
and $\Gamma=\left[\begin{smallmatrix} 1 &-1 \\-1 & 1\end{smallmatrix}\right]$
are depicted in Figure 2. The coupling weights
$c_{ij}$ associated with the edges in this case are shown in Figure 3.
To illustrate Theorem 2, let the communication graph be
$\mathcal {G}_1$ in Figure 1(a) and $\tau_i=1$, $i=1,\cdots,6$, in \dref{clc4}.
The consensus errors $x_i-x_1$, $i=2,\cdots,5$, of the third-order integrators
under the protocol \dref{clc4} with $F$, $L$, and $\Gamma$ as above
are depicted in Figure 4. The coupling weights
$d_{i}$ associated with the nodes are drawn in Figure 5.
Figures 2 and 4 state that consensus is indeed
achieved in both cases. From Figures 3 and 5, it can be observed that
the coupling weights
converge to finite steady-state values.

\begin{figure}[htbp]\centering
\centering
\includegraphics[height=0.26\linewidth,width=0.4\linewidth]{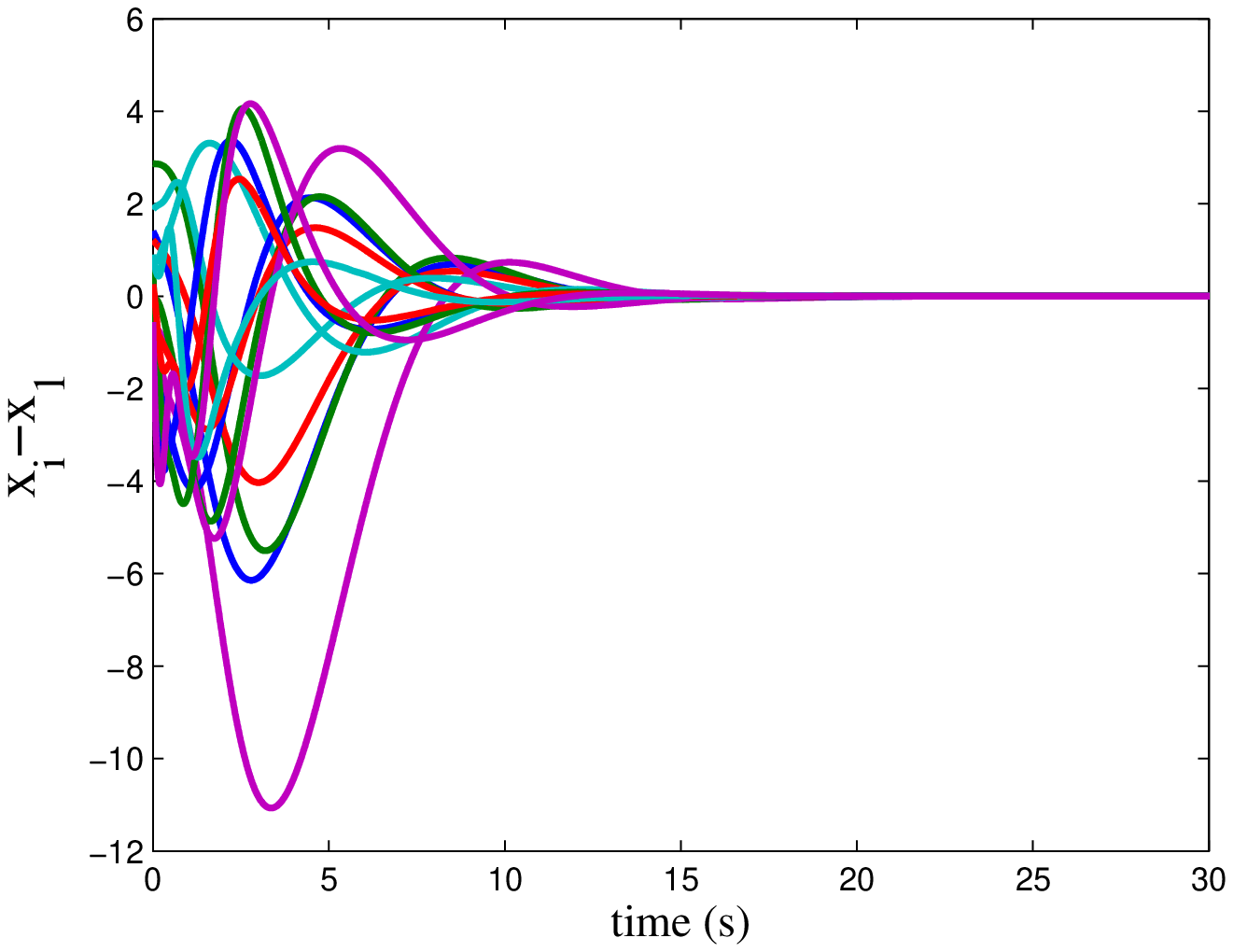}
\caption{The consensus errors $x_i-x_1$ of third-order integrators under \dref{clcsw}.}
\end{figure}

\begin{figure}[htbp]\centering
\centering
\includegraphics[height=0.26\linewidth,width=0.4\linewidth]{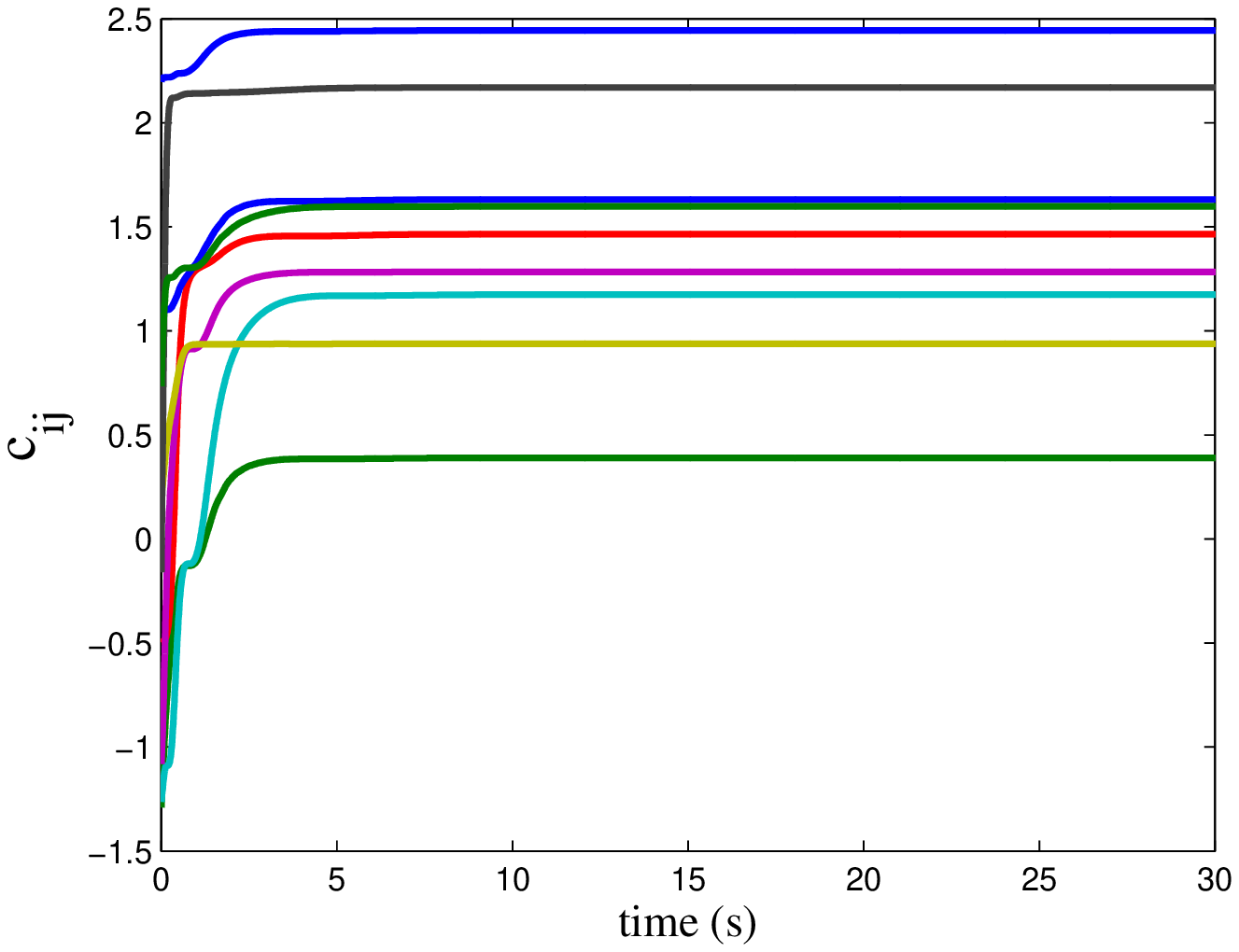}
\caption{The coupling weights $c_{ij}$ associated with the edges in \dref{clcsw}.}
\end{figure}

\begin{figure}[htbp]\centering
\centering
\includegraphics[height=0.26\linewidth,width=0.4\linewidth]{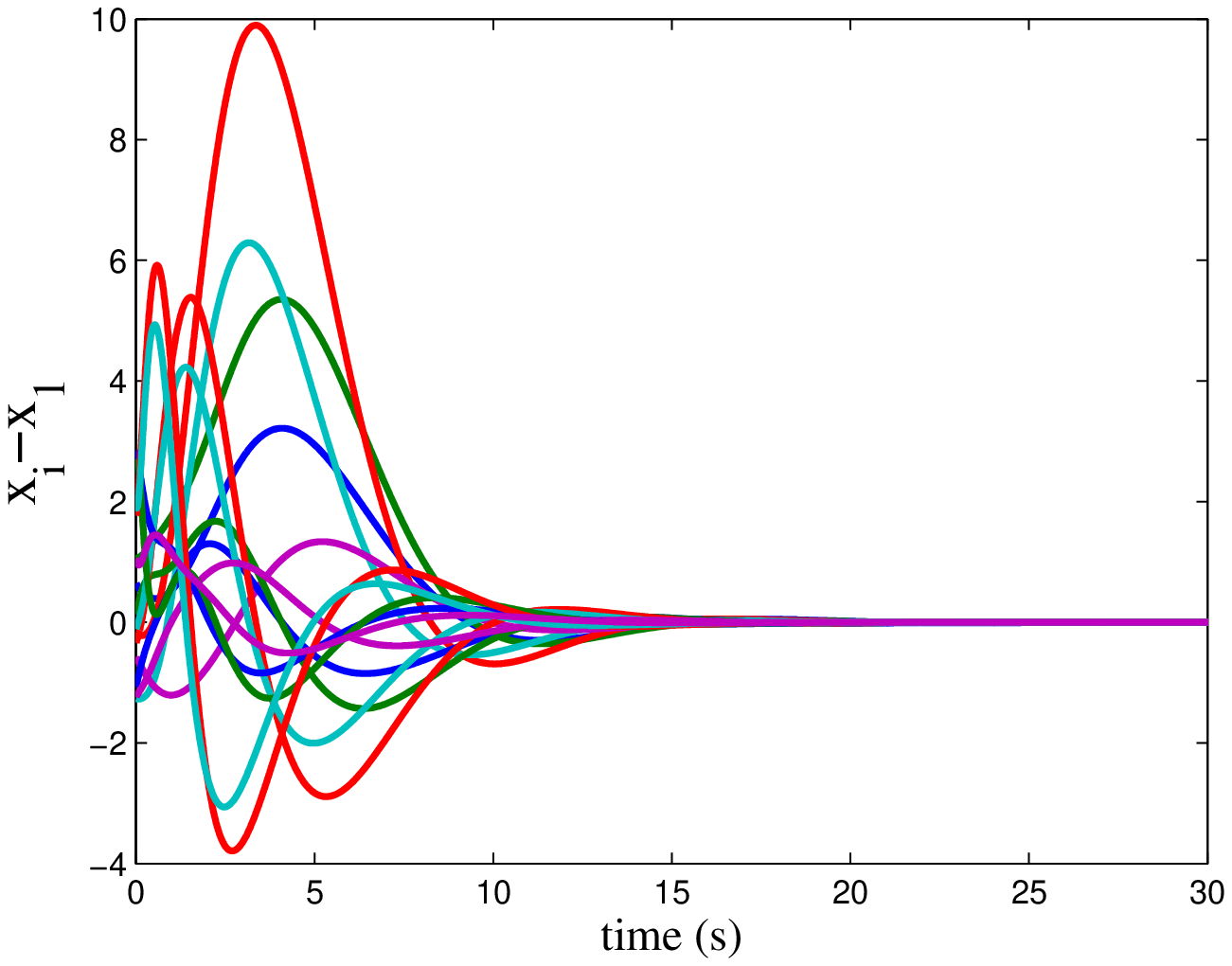}
\caption{The consensus errors $x_i-x_1$ of third-order integrators under \dref{clc4}.}
\end{figure}

\begin{figure}[htbp]\centering
\centering
\includegraphics[height=0.26\linewidth,width=0.4\linewidth]{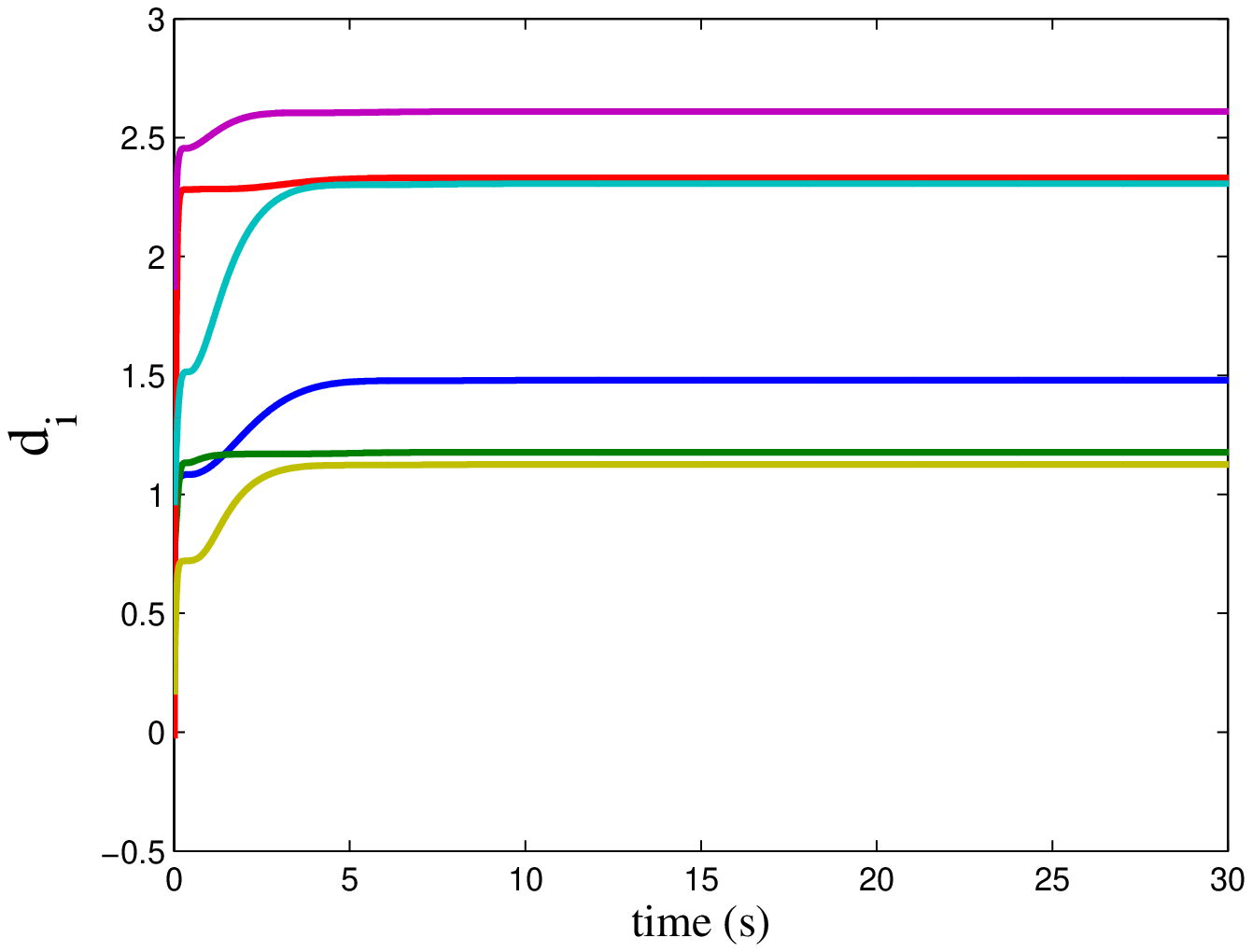}
\caption{The coupling weights $d_{i}$ associated with the nodes in \dref{clc4}.}
\end{figure}

\section{Conclusion}


In this paper, the consensus problem
of multi-agent systems with identical general
linear dynamics has been considered.
Based on the relative
output information of neighboring agents,
two distributed adaptive dynamic consensus protocols have been proposed,
namely, one protocol assigns
an adaptive coupling weight to each edge in the communication graph
while the other uses an adaptive coupling weight for each node.
These two adaptive protocols have been designed to ensure that
consensus is reached in a fully distributed fashion for
any undirected connected communication graphs without using any
global information.

%
%

%
%

\end{document}